 \documentclass[twocolumn,showpacs,preprintnumbers,amsmath,amssymb]{revtex4} \usepackage{./figures}

\usepackage{./commands}
\usepackage{./biblio_aps}

\renewcommand{\acknowledgments}{\section*{\uppercase{acknowledgments}}}
\newcommand{\linlin}[1]{\mbox{#1D-lin$\perp$lin}}

\begin{document}

\title{Synchronization of Hamiltonian motion and dissipative effects \\ in optical lattices:
        Evidence for a stochastic resonance}

\author{Laurent Sanchez-Palencia}
  \email{lsanchez@lkb.ens.fr}
  \homepage{http://www.lkb.ens.fr/~lsanchez}
\author{Gilbert Grynberg}
\affiliation{Laboratoire Kastler Brossel, D\'epartement de Physique de l'Ecole
Normale Sup\'erieure, 24, rue Lhomond, 75231, Paris Cedex 05,
France.}

\date{\today}

\begin{abstract}
We theoretically study the influence of the noise strength on the excitation of the Brillouin propagation modes in a dissipative optical lattice. We show that
the excitation has a resonant behavior for a specific amount of noise corresponding to the precise synchronization of the Hamiltonian motion on the optical
potential surfaces and the dissipative effects associated with optical pumping in the lattice. This corresponds to the phenomenon of stochastic resonance.
Our results are obtained by numerical simulations and correspond to the analysis of microscopic quantities (atomic spatial distributions) as well as macroscopic
quantities (enhancement of spatial diffusion and pump-probe spectra). We also present a simple analytical model in excellent agreement with the simulations.
\end{abstract}

\pacs{05.45.-a,05.60.-k,32.80.Pj}

\maketitle

\section{Introduction}
\label{introduction}
During the past decades, the rapid development of the experimental cooling and trapping methods for  neutral particles
has delivered a remarkable mastery of accurate measurement procedures of
the dynamical and spectral properties of the atoms \cite{nobel97}. Cold atomic samples thus constitute
privileged media for the study of various physical phenomena
such as the properties of degenerate quantum gases of both bosons \cite{nobel01} and fermions
\cite{fermions}, or nonlinear mechanical effects \cite{nonlinear}.
Among the cooling schemes, the Sisyphus mechanism \cite{sisyphus89} distinguishes by
the peculiarity that it leads not only to very cold atomic samples but also to the trapping of the atoms in
spatially periodic potential wells \cite{castin91,verkerk92,localization} by pure optical means. This creates a periodic array of atoms
bound by light, the so-called {\it optical lattices}. Moreover, the trapping potentials
can be easily designed in a large variety of different topographies in one, two or three dimensions \cite{jessen96,grynberg01}
and this provides the possibility of modeling different situations. The cooling mechanism itself
results from the intercombination of Hamiltonian motion and dissipative processes and is thus particularly well
suited for the study of the dynamics of particles trapped in periodic structures in the presence of noise.

In the Sisyphus scheme, a two level atom with Zeeman degeneracy
interacts with a laser field with a gradient of the polarization. The light induces a spatially periodic modulation of the
optical potentials (light shifts) of the different Zeeman sublevels and also random quantum jumps between
these sublevels (optical pumping). The cooling mechanism results from successive cycles of
Hamiltonian motion on the optical potential surfaces followed by dissipative processes associated to the random
jumps in such a way that a moving atom on average climbs up potential hills before it is optically pumped into a
lower lying potential surface \cite{sisyphus89}. The kinetic energy is thus converted into potential energy which
is subsequently carried away by a spontaneously emitted photon, thereby reducing the total atomic energy.
Because the steady-state kinetic energy is generally lower than the potential
depth, the particles get trapped in the potential wells. The typical evolution times corresponding to the Hamiltonian
motion on the one hand and to the dissipative processes on the other hand can vary independently with the
lattice parameters and for usual experimental conditions have roughly the same order of magnitude. This
leads to distinguish two dynamical regimes (the {\it jumping regime} where dissipative effects dominate and the
{\it oscillating regime} where Hamiltonian motion dominates). This makes particularly difficult to perform analytical
calculations in the whole accessible range of parameters \cite{houches90}.

By contrast, the intermediate regime offers the possibility of studying unusual and fascinating
atomic behaviors combining Hamiltonian motion and dissipative effects.
In particular, it can be considered as models for the transport of particles trapped in periodic structures and in
the presence of noise, which is of particular interest for systems where the dynamics is dominated
by noise such as in many biological systems \cite{biology}.
For example, an {\it atomic ratchet} has been realized in a variant of the configuration considered in this
paper \cite{ratchet} and the phenomenon of {\it directed diffusion} has been observed \cite{schiavoni03}
in a one-dimensional setup.

In a recent publication \cite{stochastic1}, we identified a resonant behavior of the stimulated transport
({\it Brillouin propagation modes}) of atoms versus the noise strength, {\it i.e.} versus the rate of optical pumping, and we
interpreted it in terms of a {\it stochastic resonance} (SR). This phenomenon, corresponding to a situation where Hamiltonian motion
and noise-induced effects work together to induce a resonant response of a system to a weak external stimulation, has been widely
studied in various theoretical models \cite{stochastic}.
However, no direct evidence for SR in dissipative optical lattices has been given so far.
The aim of this paper is to theoretically analyze the influence of the level of noise on the efficiency of the excitation process
of the Brillouin propagation modes in a dissipative optical lattice. We show that the excitation is resonant for a specific amount
of noise corresponding precisely to the synchronization of the hamitonian motion and of the dissipative effects. Our
results are mainly obtained by means of numerical simulations.

Our paper is organized as follows.
In section~\ref{dynamics}, we present the main features of the dynamics of atoms cooled and trapped in a dissipative
optical lattice. We discuss the peculiarities of Hamiltonian motion and dissipative processes and we describe the Brillouin
propagation modes.
In the next two sections, we show that the excitation of the propagation modes is resonant for a given amount of noise and
that this is compatible with the phenomenon of SR: (i) in section~\ref{macroscopic}, we analyze the resonant
enhancement of the spatial diffusion of the atomic cloud and we show that the position of the resonance precisely corresponds to
the synchronization of the Hamiltonian motion and the dissipative processes;
(ii) in section~\ref{microscopic}, we present further evidences for the phenomenon of SR by considering the
microscopic behavior of the propagating atoms. We also show that SR can be observed in the amplitude of the
Brillouin resonance that appears in well-mastered nonlinear spectroscopy experiments.
Finally, we summarize our results in section~\ref{conclusion}.

\section{Hamiltonian motion and dissipative processes in optical lattices}
\label{dynamics}

\subsection{Dynamics of cold atoms in an optical lattice}
\label{cold}
Consider a sample of atoms interacting with a laser field with a frequency slightly red detuned with respect to a
\eqt{J \rightarrow J+1} atomic transition.
The atoms experience a set of optical potentials corresponding to the
light shifts of the various Zeeman sublevels of the atomic ground state and thus depending on the internal state.
The topography of the potential surfaces depends on the interference pattern of the laser fields and are spatially periodic
with periods of the order of the laser wavelength.
In this paper we study the so-called \linlin3 configuration \cite{3dlinlin}. This is obtained from the standard \linlin1
configuration \cite{sisyphus89} by symmetrically splitting each of the two laser beams into two parts at an angle
\eqt{\theta} with the \eqt{(Oz)} axis in the \eqt{(Oxz)} and \eqt{(Oyz)} planes
respectively. The resulting configuration consists of two pairs of laser beams in the \eqt{(Oxz)} plane and in the
\eqt{(Oyz)} plane, respectively with orthogonal linear polarizations as depicted in Fig.~\ref{fig:tetra}.
In this situation, the potential surfaces are made of regularly displaced wells (see Fig.~\ref{fig:brillouin}).

\bfig
\infig{25em}{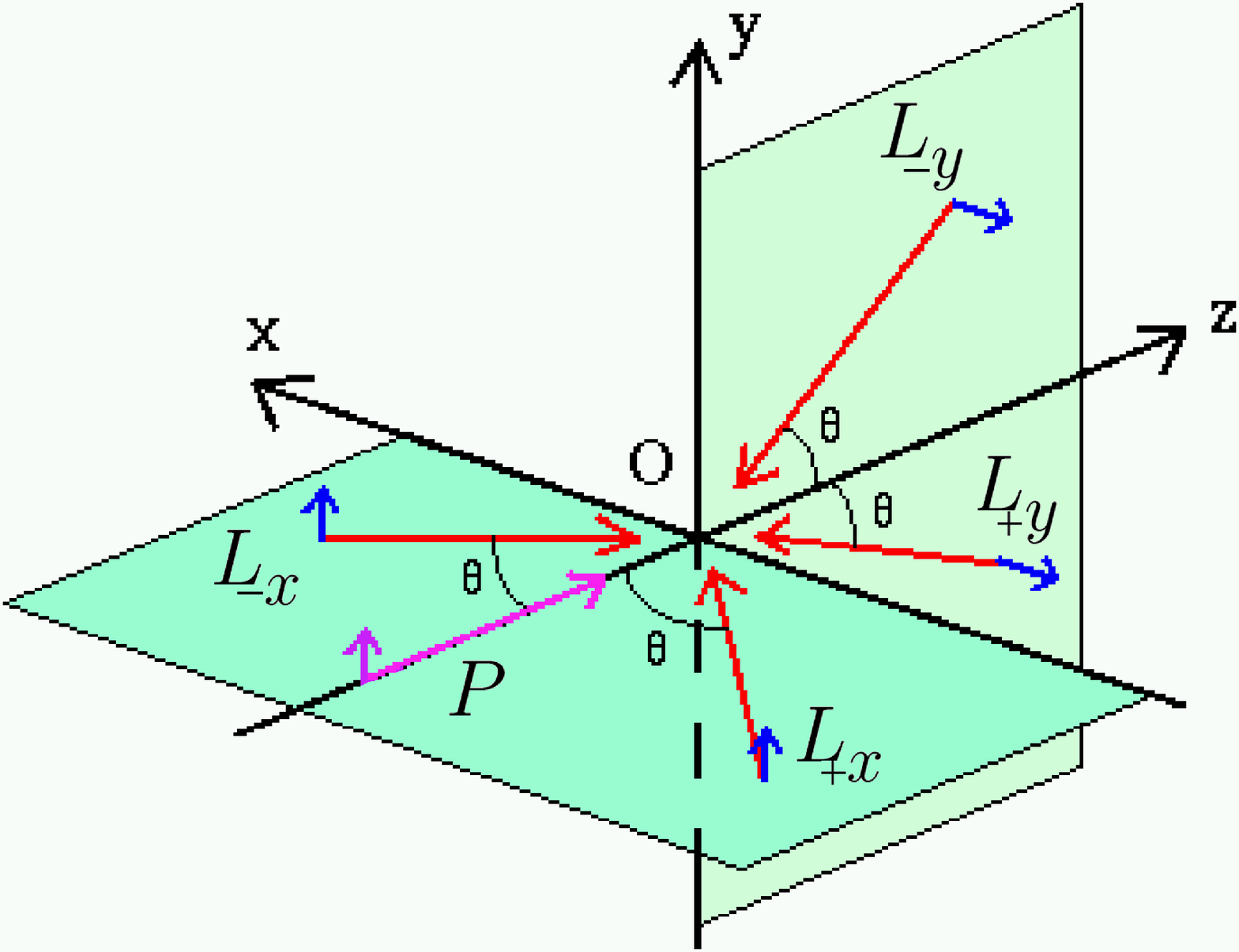}
\caption{\linlin3 laser configuration: the interference pattern of the four laser beams \eqt{L_{\pm \mu}} (with \eqt{\mu \in \{x, y\}}), with
linear polarizations arranged as in the figure, creates the optical lattice.
A probe laser beam (\eqt{P}) along the \eqt{z}- axis is also added for excitation of propagation modes (see section~\ref{propagation}) and
nonlinear spectroscopy (see section~\ref{spectroscopy}).}
\label{fig:tetra}
\efig

Moreover, for a near resonant laser field, the optical pumping cycles transfer the atoms from a given sublevel
to another one with a rate depending on the atomic position. More precisely, the optical
pumping tends to populate the lower potential surfaces thus providing the cooling of the moving atoms via the
Sisyphus effect \cite{sisyphus89,houches90}. The resulting mean atomic kinetic energy is lower than the
potential barrier for escaping from the potential wells, so that the atoms get trapped and form a crystal-like lattice
bound by light.

After the cooling stage, the atomic dynamics is composed of two parts: the Hamiltonian motion and dissipative effects.
First, the hamitonian motion
consists in periodic oscillations around the bottom of potential wells. The principal directions of the oscillations
correspond to the coordinates axes \eqt{x}, \eqt{y} and \eqt{z} with a frequency \eqt{\Omega_{\mu}/2\pi} along
the \eqt{\mu}- axis. In particular, in the \linlin3 configuration and for a
\eqt{J_\textrm{g}=1/2 \rightarrow J_\textrm{e}=3/2} atomic transition, one finds
\be
\Omega_x = 4 \sin\left( \theta \right) \sqrt{|\Delta_0'| \omega_\textrm{r}}
\label{eq:omegax}
\ee
where \eqt{\theta} is half the angle between the quasi-copropagating laser beams (see Fig.~\ref{fig:tetra}) and
\eqt{\Delta_0'} is the light shift per lattice beam and for a Clebsch-Gordan coefficient of \eqt{1} \cite{grynberg01}.
\eqt{\omega_\textrm{r} = \frac{\hbar k^2}{2M}}, with \eqt{k}, the laser wavenumber and \eqt{M} the mass of one atom, denotes the recoil
frequency. In the following, we only consider the case \eqt{\theta = 30\degree}.
Second, the dissipative processes correspond to random quantum jumps between the various potential surfaces
that modify instantaneously the atomic energy. For the same conditions as for Eq.~(\ref{eq:omegax}) and in the semi-classical
approximation, the optical pumping rate from the internal state \eqt{\left| g, \pm \right\rangle} to the state \eqt{\left| g, \mp \right\rangle}
averaged over position, reads \cite{castin94}
\be
\gamma_{\pm,\mp} \simeq \frac{4 \Gamma_0'}{9} \left\{ \overline{\cos^2 (2\pi \hat{x}) + \cos^2 (2\pi \hat{y})} \right\}
\label{eq:escape}
\ee
where \eqt{\Gamma_0'} is the scattering rate of photons per lattice beam and for a Clebsch-Gordan
coefficient of \eqt{1}, \eqt{\hat{x}} and \eqt{\hat{y}} denotes the atomic position along the \eqt{x}- and \eqt{y}- directions
in units of the lattice spacing in the corresponding direction and \eqt{\overline{(~.~)}} denotes the uniform spatial average
value \footnote{Here, \eqt{\hat{\mu} = \mu / \lambda_{\mu}}, where \eqt{\lambda_{\mu}} is the lattice spacing in direction \eqt{\mu \in \{x,y,z\}}.
For the \linlin3 configuration, \eqt{\lambda_{x,y} = \lambda / (k \sin(\theta))} and \eqt{\lambda_{z} = \lambda / (2 k \cos(\theta))}.}.

As pointed out in section~\ref{introduction}, the relative importance of Hamiltonian evolution and dissipative effects depends on the lattice parameters.
In the jumping regime (\eqt{\Gamma_0' \gg \Omega_x}),
the dissipative effects dominate the dynamics so that an atom encounters several optical pumping cycles during a
single oscillation in a potential well.
On the contrary, in the oscillating regime (\eqt{\Omega_x \gg \Gamma_0'}), the dynamics is dominated by the
Hamiltonian motion and an atom makes several oscillations before being optically pumped.
Note that in our configuration, \eqt{\Delta_0' \propto I_\textrm{L}/\Delta} and \eqt{\Gamma_0' \propto I_\textrm{L}/\Delta^2}, where
\eqt{I_\textrm{L}} is the intensity of one laser beam and \eqt{\Delta} is the detuning between the laser and the Bohr
frequencies of the atomic transition, so that the typical times of both Hamiltonian motion (\eqt{2\pi/\Omega_x})
and of the dissipative effects (\eqt{1/\gamma_{\pm,\mp}}) can vary independently in a large range.

Finally, because a strong trapping site in the potential surface for a given internal state (for example
\eqt{x=y=z=0} for \eqt{\left|-\right\rangle}) corresponds to a non-trapping site
in the potential surface for an other internal state (see Fig.~\ref{fig:brillouin}), \eqt{\gamma_{\pm,\mp}} is approximately the escape
rate from a trapping site. Note that the macroscopic manifestation of the escapes from trapping sites consists in spatial diffusion of
the atomic cloud in the lattice \cite{sanchez02}.

\subsection{Brillouin Propagation modes}
\label{propagation}
One can excite atomic propagation modes in the optical lattice by adding a laser beam (\eqt{P}) detuned by \eqt{\delta} from the lattice
beams and aligned with the symmetry axis \eqt{z} as shown in Fig.~\ref{fig:tetra} \cite{courtois96,stochastic1}. This laser is usually
referred to as the {\it probe beam} because it was first introduced in nonlinear spectroscopy experiments in optical lattices
(\cite{verkerk92,hemmerich93}, see also section~\ref{spectroscopy}).
For an atom, a propagation mode in a given direction consists in making a half oscillation in a
potential well followed by an optical pumping cycle that changes its internal state, then in making a new half oscillation in the
neighboring well, undergoing an optical pumping cycle and so on, as shown in Fig.~\ref{fig:brillouin}.

\bfig
\infig{25em}{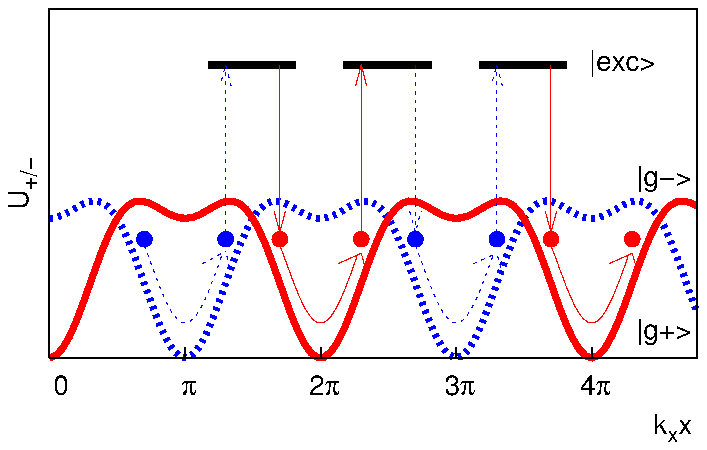}
\caption{Cut along a given direction (\eqt{x}) of the optical potentials \eqt{U_\pm} for a
\eqt{J_\textrm{g}=1/2 \rightarrow J_\textrm{e}=3/2} atomic transition.
The propagation mode along \eqt{+x} is also represented.
Similar modes exist along \eqt{-x}, \eqt{\pm y} and \eqt{\pm z}.}
\label{fig:brillouin}
\efig

The interference between the probe laser beam and the lattice beams induces a moving spatial modulation of the potential
surfaces. For the configuration considered here, there are two modulations symmetric with respect to \eqt{z} along the directions
\eqt{\vect{u}^\pm = \vect{\Delta k}^\pm / |\vect{\Delta k}^\pm|} where \eqt{\vect{\Delta k}^\pm} is the difference between the wavevectors of
\eqt{L_{\pm x}} and \eqt{P} and that lies in the \eqt{(Oxz)} plane.
Note that there are also two moving potential modulations in the \eqt{(Oyz)} plane. However, we do not consider them because they are of a very different
kind and in particular, they excite the propagation modes in the \eqt{(Oyz)} plane only for a value of \eqt{\delta} different from the one considered in the
following (see Eq.~(\ref{eq:brillouin})).
The potential modulations are periodic in space with a period
\eqt{\lambda_\textrm{mod} = 2\pi / |\vect{\Delta k}^\pm|} and move along \eqt{\vect{u}^\pm} at a velocity
\eqt{v_\textrm{mod} = \delta / |\vect{\Delta k}^\pm|}. In particular, in the \eqt{x}- direction, the modulations move along \eqt{+x} and \eqt{-x}
at a velocity \eqt{v_x = \delta / |\vect{\Delta k}^\pm \cdot \vect{e_x}|}. Because the atoms tend to populate the potential wells of the
moving modulation, the two modulations dragg the atoms along \eqt{+x} and \eqt{-x} at a velocity \eqt{v_\textrm{mod}}.
Hence, a resonant excitation ({\it Brillouin resonance}) of the propagation mode is obtained when the phase velocity
\eqt{v_\textrm{mod}} of the driving modulations is equal to the natural velocity \eqt{v_x} of an atom along the mode in the lattice. \eqt{v_x}
corresponds (see Fig.~\ref{fig:brillouin}) for an atom to travel over one half spatial period (\eqt{\pi / k_x}) in half a temporal oscillating period
\eqt{\pi / \Omega_x}, so that \eqt{v_x = \Omega_x / k_x}. In the \linlin3 lattice, one finds that the Brillouin resonance for the propagation
modes along the \eqt{x}- direction corresponds
to
\be
\delta = \pm \Omega_x~.
\label{eq:brillouin}
\ee

In the following, we consider situations where the propagation modes in the \eqt{x}- direction are resonantly excited versus \eqt{\delta}.
The atomic cloud thus separates into three modes.
The first one corresponds to atoms trapped in the potential wells of the unmoving lattice and that are not sensitive to the moving potential
modulations whereas the other two correspond to atoms moving along the propagation modes at a velocity \eqt{v_\textrm{mod}} towards \eqt{+x}
or \eqt{-x}. These modes are not independent. There are population transfers between the modes, mainly due to non-Hamiltonian
forces encountered by an atom which are related to random optical pumping cycles and fluorescence cycles.

\subsection{Numerical simulations}
\label{simulation}
Our theoretical analysis is supported by numerical simulations. These consist in a Monte-Carlo integration of the master equation for the
dynamics of an atom interacting with a laser field and the vacuum modes in the semi-classical approximation \cite{castin94,sanchez02}.
This approximation is justified in the range of parameters that we consider here because of the strong localization of the atomic wavefunctions
\cite{houches90}. We consider a theoretical transition \eqt{J_\textrm{g}=1/2 \rightarrow J_\textrm{e}=3/2} which gives good qualitative
results and for which analytical calculations can be performed \cite{sisyphus89}.

In the semi-classical approximation of the laser cooling theory, the master equation can be written as a Fokker-Planck equation (FPE)
\cite{castin94,marksteiner96,sanchez02} where the external degrees of freedom of the atoms are treated as classical variables.
This is obtained from the Wigner transform \cite{wigner32} of the full quantum master equation for external as well as internal degrees
of freedom under the assumption of a momentum distribution which is much broader than a single photon momentum, \eqt{\Delta P\gg \hbar k}.
The FPE for the populations \eqt{\Pi_\pm({\bf r}, {\bf p}, t)} of \eqt{|\pm\rangle} reads
\bea
& &\left[\partial_t + \frac{p_i}{M} \partial_i - (\partial_i U_\pm) \partial_{p_i} \right] \Pi_\pm = \nonumber \\
& & \quad\quad \gamma_{\mp,\pm} \Pi_\mp - \gamma_{\pm,\mp} \Pi_\pm  \nonumber \\
& & \quad\quad - F^i_{\pm\pm} \partial_{p_i} \Pi_\pm - F^i_{\mp\pm} \partial_{p_i} \Pi_\mp \nonumber \\
& & \quad\quad + D^{ij}_{\pm\pm} \partial_{p_i} \partial_{p_j} \Pi_\pm + D^{ij}_{\mp\pm} \partial_{p_i} \partial_{p_j} \Pi_\mp.
\label{eq:FPE}
\eea
Here \eqt{i,j=x,y,z} and summation over \eqt{i} and \eqt{j} is assumed. In this equation, \eqt{\gamma_{\pm,\mp}} is the jumping rate
from the Zeeman sublevel \eqt{|\pm\rangle} to the sublevel \eqt{|\mp\rangle}, \eqt{F^i_{\pm\pm}} represents the radiation pressure
force and \eqt{D^{ij}_{\pm\pm}} the momentum diffusion matrix for atoms in the internal state \eqt{|\pm\rangle}. \eqt{F^i_{\pm\mp}}
and \eqt{D^{ij}_{\pm\mp}} are the corresponding quantities associated with jumps between different internal states \cite{petsas99}.
Note that all these coefficients are laser field depend. In our case the laser field is made of the contributions of the lattice beams \eqt{L_{\pm \mu}}
and of the probe beam \eqt{P}.
A numerical solution of the FPE can be obtained by averaging over many realizations of the corresponding Lan\-ge\-vin equations
\cite{risken89,sanchez02}.

For the sake of simplicity and to save computation time, we restricted
the atomic motion into two dimensions of space, in the plane \eqt{y = 0}. That way, we consider the motion along \eqt{x} (direction of the stimulated propagation
modes) and also along \eqt{z} to take into account the possible escapes in the orthogonal direction. As pointed out in the preceeding section, we only
consider cases where only the propagation modes in the \eqt{(Oxz)} are excited. We therefore expect our results to differ from 3D situations only by
scaling factors due to the atomic spatial distribution in the \eqt{y} direction (see also the discussion at the end of section~\ref{characterization}).

\section{Macroscopic manifestation of stochastic resonance}
\label{macroscopic}

\subsection{Resonant enhancement of spatial diffusion}
\label{enhancement}
The excitation of the propagation modes drives one part of the atomic cloud along \eqt{+x} and another
part along \eqt{-x}. The spatial diffusion in the specific direction \eqt{x} thus results enhanced whereas
the spatial diffusion in the orthogonal directions is almost unchanged \cite{stochastic1}. In order to optimize
this phenomenon for a given set of lattice parameters, we  tune the probe laser frequency to the Brillouin
resonance according to Eq.~(\ref{eq:brillouin}).
Because the propagation process of an atom along a mode involves both Hamiltonian motion and dissipative
processes, it is clear that changing the strength of one with respect to the other should modify the efficiency of
the driving.
We plot in Fig.~\ref{fig:stocdirect} the spatial diffusion coefficient in the lattice along the driving
direction (\eqt{D_x}) and in the orthogonal direction (\eqt{D_z}) as a function of the noise strength (proportional
to the optical pumping rate \eqt{\Gamma_0'}).

\bfig
\infig{25em}{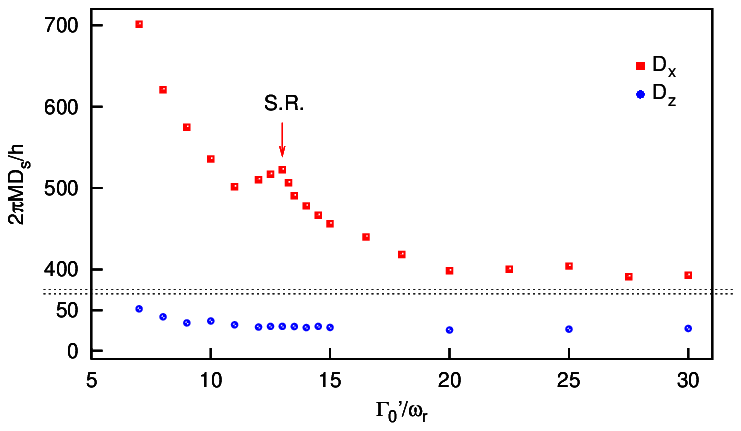}
\caption{
Numerical results for the spatial diffusion coefficients in the \eqt{x}- and \eqt{z}- directions as
a function of the optical pumping rate, for a given depth of the optical potential.
Parameters for the calculations are: \eqt{\Delta_0' = -200 \omega_\textrm{r}} and
\eqt{\theta=30\degree}.}
\label{fig:stocdirect}
\efig

Generically, \eqt{D_{\mu}} shows a monotonous decrease versus \eqt{\Gamma_0'}. This behavior is in agreement
with a simple model of random walk \cite{itzykson91}: an atom is trapped in a potential well with a lifetime of the order of
\eqt{\tau_\textrm{esc} \sim 1/\Gamma_0'} and during a jump it travels at a typical velocity of
\eqt{\overline{v}=\sqrt{2 E_\textrm{K} / M}} (where \eqt{E_\textrm{K}} is the kinetic energy of the cloud and \eqt{M} is the mass of one atom)
for a typical time \eqt{\tau_\textrm{flight} \sim 1/\Gamma_0'}. Because for Sisyphus cooling,
\eqt{E_\textrm{K} \propto \hbar|\Delta_0'|} \cite{sisyphus89}, the spatial diffusion coefficient scales as
\eqt{D_{\mu} \sim \wee{\hbar}{M}\wee{|\Delta_0'|}{\Gamma_0'}} \cite{sanchez02}.
In addition to the global behavior versus \eqt{\Gamma_0'}, we find that \eqt{D_x} displays a clear
narrow peak for a specific value of \eqt{\Gamma_0'} (\eqt{\Gamma_0' \simeq 13 \omega_\textrm{r}} in
Fig.~\ref{fig:stocdirect}) whereas no peak is observed for the spatial diffusion coefficient along the
orthogonal (\eqt{z}-) direction. This corresponds to a resonant enhancement of the spatial diffusion along
the driving direction and can be interpreted as a resonant excitation of the Brillouin propagation modes for a given
amount of noise. This point will be demonstrated in detail in section~\ref{bunching}.

\subsection{Characterization of stochastic resonance}
\label{characterization}
In order to characterize the specific influence of the propagation modes to the spatial diffusion in the \eqt{x}-
direction, we introduce the enhancement factor of the spatial diffusion coefficient along \eqt{x} with a probe
beam at Brillouin resonance (see Eq.~(\ref{eq:brillouin})) \eqt{D_x} compared to far from Brillouin
resonance \eqt{D_x^0}:
\be
\xi = \frac{D_x - D_x^0}{D_x^0}~.
\label{eq:enhancement}
\ee
Here, \eqt{D_x^0} is typically calculated for \eqt{\delta \simeq 100 \Omega_x}.
In Fig.~\ref{fig:stoc2d}, we plot the enhancement factor \eqt{\xi} versus the noise strength \eqt{\Gamma_0'}
for a fixed value of the potential depth \eqt{\Delta_0'} and for various depths of the driving modulation induced
by the probe beam.

\bfig
\infig{25em}{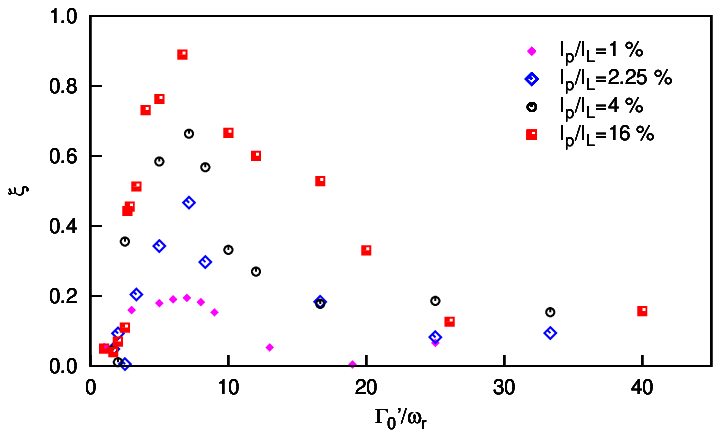}
\caption{
Numerical results for the enhancement factor \eqt{\xi} as a function of the optical pumping rate, for a given depth of
the optical potential. The four data sets correspond to different intensities of the probe beam \eqt{I_\textrm{P}} in units of the
intensity of one lattice beam \eqt{I_\textrm{L}}, {\it i.e.} different depths of the moving potential modulation. Parameters for the
calculations are: \eqt{\Delta_0' = -50 \omega_\textrm{r}} and \eqt{\theta=30\degree}.}
\label{fig:stoc2d}
\efig

The results of Fig.~\ref{fig:stoc2d} show that \eqt{\xi} is a nonmonotonic function of the amount of
noise \eqt{\Gamma_0'} in the lattice. For low noise strength, \eqt{\xi} is an increasing function of
\eqt{\Gamma_0'}, then \eqt{\xi} reaches a maximum and finally decreases
for high values of the noise strength \eqt{\Gamma_0'}. Hence, there is a specific value of the noise strength
for which the propagation modes are resonantly excited. Such a property is interpreted as a SR and
corresponds to the synchronization of Hamiltonian motion and dissipative processes \cite{stochastic}.
We checked that the peak for \eqt{\xi} precisely corresponds to the peak for \eqt{D_x} and we conclude that
the resonant enhancement of spatial diffusion is due to a resonant excitation of the propagation modes in the \eqt{x}-
direction.

In order to understand more precisely the phenomenon in our system, consider again the propagation
process of an atom in a Brillouin mode (section~\ref{propagation} and Fig.~\ref{fig:brillouin}).
Such a process clearly works better when an optical pumping cycle occurs each time the atom gets close to
a crossing point between the potential curves \eqt{U_+} and \eqt{U_-}. Taking into account the velocity of the
atoms in the propagation modes \eqt{v_x=\Omega_x/k_x}, this occurs periodically in time every
\eqt{\pi/\Omega_x}. Besides, the typical time between two subsequent optical pumping cycles is
\eqt{1/\gamma_{\pm,\mp}}. Hence, we expect SR to occur for \eqt{\pi/\Omega_x \simeq 1/\gamma_{\pm,\mp}},
and, according Eqs.~(\ref{eq:omegax}) and (\ref{eq:escape}), for
\be
\left(\Gamma_0'\right)_\textrm{SR} = \frac{9 \sin(\theta) \sqrt{\left|\Delta_0'\right|\omega_\textrm{r}}}
                                                        {\pi \left(\overline{\cos^2 (2\pi \hat{x}) + \cos^2 (2\pi \hat{y})}\right)}~.
\label{eq:resostoch}
\ee
For values of the noise strength lower than \eqt{\left(\Gamma_0'\right)_\textrm{SR}}, the optical pumping
is too weak to allow the atom to go across the potential barrier every time it reaches a crossing point
between the potential curves. On the contrary, for a level of noise much larger than
\eqt{\left(\Gamma_0'\right)_\textrm{SR}}, the Hamiltonian motion is broken by the numerous optical
pumping cycles so that the atoms cannot follow the driving potential modulation.

Figure~\ref{fig:stoc2d} clearly shows that the amplitude of the bell-shaped curve of \eqt{\xi} versus
\eqt{\Gamma_0'} increases for increasing depths of the driving modulation ({\it i.e.} increasing values of
\eqt{I_\textrm{P}/I_\textrm{L}}). This can be easily explained by the fact that the number of atoms dragged by the moving
modulation is greater for deeper driving modulations. We find that the position of the SR, \eqt{\left(\Gamma_0'\right)_\textrm{SR}},
is independent of \eqt{I_\textrm{P}/I_\textrm{L}} in good agreement with the property that Eq.~(\ref{eq:resostoch}) does
not involve the driving modulation properties. In fact Eq.~(\ref{eq:resostoch}) only involves the properties of
the lattice itself. Note however, that for extreme values of \eqt{I_\textrm{P}/I_\textrm{L}}, Eq.~(\ref{eq:resostoch})
certainly no longer works. Indeed, for very weak values of \eqt{I_\textrm{P}/I_\textrm{L}}, the propagation modes are not
excited and no enhancement of the spatial diffusion can be detected, whereas for very high values, the
lattice is too much perturbated to influence the dynamics of the atoms.

In order to check the validity of the theoretical synchronization condition~(\ref{eq:resostoch}), we
determined the enhancement factor \eqt{\xi} versus the noise strength \eqt{\Gamma_0'} for various
fixed values of the lattice potential depth \eqt{\Delta_0'}. We found results as in Fig.~\ref{fig:stoc2d}
and we consequently determined the position of the SR, \eqt{\left(\Gamma_0'\right)_\textrm{SR}}, as a function
of \eqt{\Delta_0'} in a wide range. We plot our
results in Fig.~\ref{fig:position2d} together with the theoretical curve corresponding to
Eq.~(\ref{eq:resostoch}) for a 2D lattice (\eqt{\hat{y}=0}) where
\eqt{\overline{\cos^2 (2\pi \hat{x}) + \cos^2 (2\pi \hat{y})}=3/2} as in the simulations. Note that no free parameter
is used to fit the numerical data.

\bfig
\infig{25em}{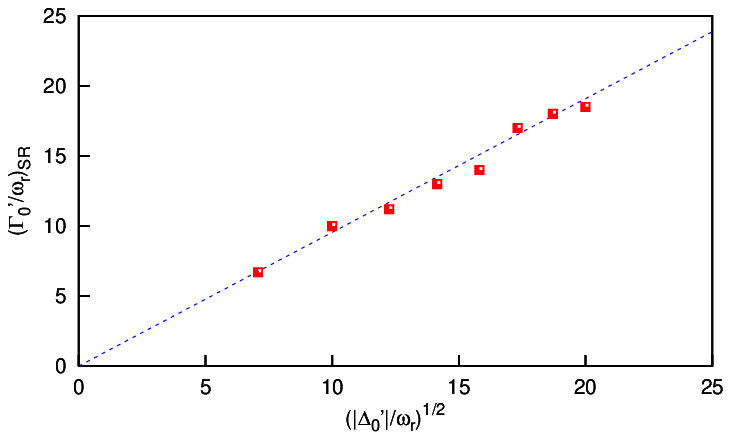}
\caption{
Position of the stochastic resonance versus the square root of the modulation
depth and comparison to the theoretical value given by Eq.~(\ref{eq:resostoch}). Parameters
for the simulations are \eqt{\theta=30\degree} and \eqt{I_\textrm{P}/I_\textrm{L}=9\%}.}
\label{fig:position2d}
\efig

We obtain an excellent qualitative as well as quantitative agreement  between the numerical results and the theory. We thus
conclude that the resonance observed in Fig.~\ref{fig:stoc2d} corresponds to the phenomenon of SR for which the
dissipative processes (optical pumping) are synchronized with the Hamiltonian motion (half-oscillations in the potential
wells) of the atoms in a propagation mode. Similar results are expected in a full 3D situation. In that case, the scaling factor of
\eqt{\left(\Gamma_0'\right)_\textrm{SR}} versus \eqt{\sqrt{|\Delta_0'|}} is however expected to be slightly different because
of a different average value for \eqt{\overline{\cos^2 (2\pi \hat{x}) + \cos^2 (2\pi \hat{y})}}. In the \eqt{x} direction, a uniform
average is valid because the atoms excited in the propagation  modes are moving along \eqt{x} but the average over \eqt{y}
should take into account the atomic localization in the lattice \cite{sanchez02}. In fact, because the potential minima correspond
to \eqt{\hat{y} \equiv 0 ~[\pi]}, we expect \eqt{\overline{\cos^2 (2\pi \hat{x}) + \cos^2 (2\pi \hat{y})}} not to be very different
from \eqt{3/2}.

\section{Microscopic dynamics and stochastic resonance}
\label{microscopic}

\subsection{Atomic bunching in a moving frame}
\label{bunching}
In section~\ref{macroscopic}, we have given evidences for the realization of SR in the enhancement of the spatial diffusion of
the atomic cloud in the \eqt{x}- direction. However, no detailed study of the microscopic behavior of the atoms in the presence of a variable amount
of noise has been performed so far. In particular, the problem arises as to give a direct evidence that the propagation modes
are really excited resonantly for a given amount of noise. In order to adress this problem, we shall directly determine the atomic population
in a propagation mode versus \eqt{\Gamma_0'}. In the configuration depicted in Fig.~\ref{fig:tetra}, two potential modulations with
a spatial period \eqt{\lambda_\textrm{mod}} and a velocity \eqt{v_\textrm{mod}}, dragg the atoms symmetrically with respect to axis \eqt{(Oz)} in the
directions \eqt{\vect{u}^+} and \eqt{\vect{u}^-}. We therefore expect that the atomic distribution is spatially modulated in the moving potential
modulations and particularly that the atoms are bunched  around the moving potential minima.

We determined the atomic spatial distribution \eqt{N^\pm (u)} in a frame moving in the direction and at the velocity of each mode. Here \eqt{u} is the
space coordinate in direction \eqt{\vect{u}^\pm}. Because of the low atomic filling rates in dissipative optical lattices, \eqt{N^\pm (u)} is in fact calculated
by accumulating over time.
Moreover, we cut the whole space in elementary \eqt{\lambda_\textrm{mod}}-large cells centered at \eqt{\lambda_\textrm{mod}(j+1/2)} and we add the populations
of each cell \eqt{N^\pm_j (u+j \lambda_\textrm{mod})}, with \eqt{u \in [0,\lambda_\textrm{mod}[}, so that
\eqt{N^\pm (u) = \sum_{j}{N^\pm_j (u+j \lambda_\textrm{mod})}}.
\eqt{N^\pm (u)} results from the superimposition of several terms corresponding to atoms with different dynamical behaviors.
On the one hand, the atoms trapped in the lattice sites and the atoms dragged along the moving potential modulation \eqt{\vect{u}^\mp} have a modulated
and stationary spatial distribution respectively in the un-moving frame and in the frame moving as the mode \eqt{\vect{u}^\mp}. Because these frames move
with respect to our reference frame (attached to the mode \eqt{\vect{u}^\pm}), their contribution is uniform with respect to \eqt{u} and contributes as a
constant number. On the other hand, the atoms dragged along the potential modulation \eqt{\vect{u}^\pm} are approximately at rest
in our frame and consequently give the shape of moving matter modulation along the mode \eqt{\vect{u}^\pm}. In a linear regime for the atomic response to
the dragging, valid for low enough values of \eqt{I_\textrm{L}/I_\textrm{P}}, we expect the contribution to \eqt{N^\pm (u)} of the atoms from mode
\eqt{\vect{u}^\pm} to be sinusoidal as the potential modulation itself. We thus conclude that the atomic spatial distribution in the moving frame should be
\be
N^\pm (u) = C \left[ 1 + A \sin \left( \frac{2 \pi u}{\lambda_\textrm{mod}} + \varphi \right) \right]~.
\label{eq:distribution}
\ee
Here the amplitude \eqt{A} characterizes the fraction of atoms in the propagation mode \eqt{\vect{u}^\pm} and the phase \eqt{\varphi} accounts for
the delay of the matter modulation with respect to the dragging potential modulation due to the finite response time of the atoms in the lattice.
\eqt{C=\int_0^{\lambda_\textrm{mod}}{\frac{du}{\lambda_\textrm{mod}} N^\pm (u)}} is determined by the total number of atoms used in the simulations and the
integration time window. \eqt{C} is thus invariant when changing the lattice parameters.

\bfig
\infig{25em}{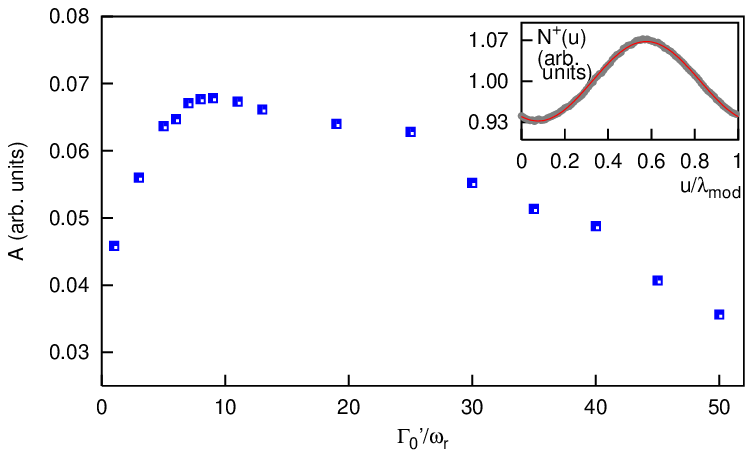}
\caption{Amplitude of the atomic fraction in the propagation mode versus the optical pumping rate. The phenomenon of stochastic
resonance is clearly visible. Parameters for the simulations are \eqt{\Delta_0' = -50 \omega_\textrm{r}}, \eqt{\theta = 30\degree} and
\eqt{I_\textrm{P} / I_\textrm{L} = 9 \%}.
Inset: Numerical result for the atomic bunching in the moving frame corresponding to the propagation mode together with a sinusoidal fit of the form
of Eq.~(\ref{eq:distribution}). Here \eqt{\Gamma_0' = 9 \omega_\textrm{r}}.}
\label{fig:bunching2d}
\efig

A typical result for \eqt{N^+ (u)} is plotted in the inset of Fig.~\ref{fig:bunching2d} together with a fit to a function of the form of Eq.~(\ref{eq:distribution})
where \eqt{A} and \eqt{\varphi} are the free parameters. The numerical results are found to be in excellent agreement with Eq.~(\ref{eq:distribution}).
We plot in Fig.~\ref{fig:bunching2d} the amplitude \eqt{A} of the sinusoidal component in \eqt{N^+ (u)} as a function of the optical pumping rate
\eqt{\Gamma_0'}. We obtain a bell-shaped curve which is characteristic of SR. This result constitutes the first direct evidence on a microscopic observable
that the Brillouin-like propagation modes are resonantly excited for a given level of dissipative effects. The position of the resonance
\eqt{\left(\Gamma_0'\right)_\textrm{SR}} is also in good agreement with the results plotted in Fig.~\ref{fig:position2d} and corresponding to the SR
observed in the enhancement factor of the spatial diffusion coefficient along the direction \eqt{x}. These results confirm by a direct method that the Brillouin
propagation modes are excited resonantly when the synchronization condition~(\ref{eq:resostoch}) between the Hamiltonian motion and the dissipative
processes in the optical lattice is fulfilled.

\subsection{Brillouin spectroscopy}
\label{spectroscopy}
A very powerful and widely used technique to probe the dynamical properties of a cold atomic sample in an optical lattice is nonlinear pump-probe
spectroscopy \cite{courtois92,verkerk92,hemmerich93} and we adress now the problem to known whether SR can be observed in spectroscopy experiments.
The method consists in using the laser (\eqt{P}) as the probe and the lattice beams (\eqt{L_{\pm \mu}}) as the pump (see Fig.~\ref{fig:tetra}). As
mentionned previously, the interference between the probe and pump beams induces moving modulations of the laser field which create time-dependent
modulations of various observables in the medium (spatial and velocity distributions, polarization, etc \dots) with a non-zero phase with respect to the laser
field modulation. This is connected to the finite response time of the atoms in the lattice. The diffraction of the pump beams onto the matter modulation
induces a modification of the probe intensity \eqt{I} which is measured as a function of the detuning between the probe and the pump beams \eqt{\delta}
\cite{courtois96_1}. The elementary excitation modes of the atomic sample appear in the spectrum (\eqt{I} as a function of \eqt{\delta}) as resonances
whose positions and widths are related respectively to the real part and to the imaginary part of the eigen-frequency of the mode \cite{spectrobooks,courtois96_2}.
It is thus clear that the amplitude of the excitation resonance in the spectrum depends on both the amplitude and the phase of the observable modulation
involved in the light scattering process. For a \linlin3 configuration with a probe beam aligned along \eqt{(Oz)}, the spectrum consists of three kinds of
resonances: first, the {\it Rayleigh lines} which relate to the pure relaxation processes associated to spatial diffusion and damping of the kinetic energy
\cite{rayleigh,carminati03}; second, the sideband {\it Raman lines} which correspond to transitions between vibrational bound states in the potential wells
\cite{courtois92}; third, the {\it Brillouin resonances} which correspond to the propagation modes discussed in section~\ref{propagation} \cite{courtois96}.

In the case of the Brillouin line, the amplitude of the resonance is closely related to the atomic spatial distribution in the moving reference frame and particularly
to the values of \eqt{A} and \eqt{\varphi}. In fact, we observed in the numerical simulations that \eqt{\varphi} weakly depends on \eqt{\Gamma_0'} for
a fixed value of \eqt{\Delta_0'} in the range of parameters considered here (the range of values for \eqt{\varphi} is found to be
\eqt{\Delta \varphi / \pi < 10\%}). Hence, the amplitude of the Brillouin resonance should characterize the amplitude \eqt{A} of the moving matter
modulation and should therefore show a SR behavior.

We performed numerical simulations of the pump-probe spectra in the lattice for different values of the optical pumping rate \eqt{\Gamma_0'} and of the
light shift per beam \eqt{\Delta_0'} by means of a method equivalent to the one described  in Ref.~\cite{courtois96_1}. In order to determine the amplitude
of the Brillouin resonance, we fit the spectra to functions of the form
\bea
I(\delta) = A_\textrm{e}\delta+B_\textrm{e}
            + A_\textrm{R} \exp\left[-\frac{(\delta-\Omega_\textrm{R})^2}{2\sigma_\textrm{R}^2}\right] + \nonumber \\
            + A_\textrm{B} \exp\left[-\frac{(\delta-\Omega_\textrm{B})^2}{2\sigma_\textrm{B}^2}\right]
\label{eq:spectrum}
\eea
with \eqt{A_\textrm{e}}, \eqt{B_\textrm{e}}, \eqt{A_\textrm{R}}, \eqt{\sigma_\textrm{R}}, \eqt{A_\textrm{B}} and \eqt{\sigma_\textrm{B}}
as free parameters. The linear term in Eq.~(\ref{eq:spectrum})
accounts for the very large central Rayleigh resonance \cite{carminati03} in a restricted range for \eqt{\delta} close to the Brillouin resonance, whereas
the two Gaussian functions account for the Raman (R) and Brillouin (B) lines centered at \eqt{\Omega_\textrm{R,B}} and with a half-width at \eqt{1/\sqrt{e}}
equal to \eqt{\sigma_\textrm{R,B}}. The {\it a priori} justification of the Gaussian shapes of the Raman and Brillouin lines is out of the scope of this paper
and the approximation of the spectra to Eq.~(\ref{eq:spectrum}) is validated by the fact that it fits well the numerical spectra as shown in the inset of
Fig.~\ref{fig:spectra2d}.

\bfig
\infig{25em}{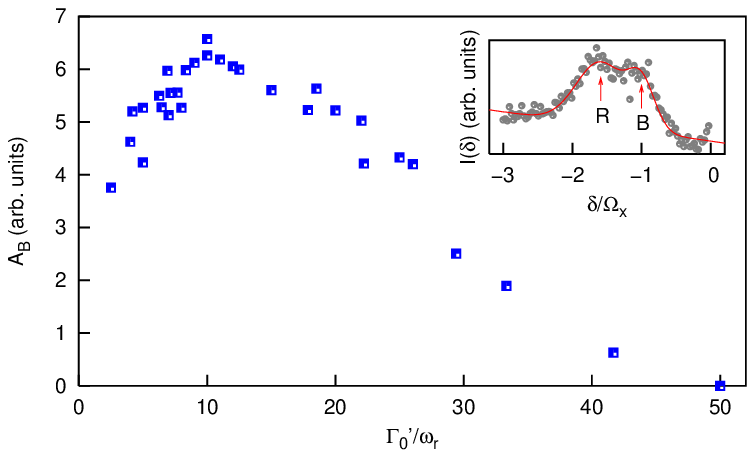}
\caption{Amplitude of the Brillouin line as a function of the optical pumping rate for a fixed potential depth as obtained from nonlinear
spectroscopy simulations. Parameters for the simulations are the same as in Fig.~\ref{fig:bunching2d}.
Inset: Numerical result of a nonlinear spectrum (\eqt{\Gamma_0' = -5.55 \omega_\textrm{r}}); the Raman (R) and Brillouin (B) resonances are
indicated on the plot.}
\label{fig:spectra2d}
\efig

We finally plot in Fig.~\ref{fig:spectra2d} the amplitude of the Brillouin resonance \eqt{A_\textrm{B}} as a function of the optical pumping rate
\eqt{\Gamma_0'} for
a fixed value of the light shift per beam \eqt{\Delta_0'}. Again, we find a bell-shaped curve which corresponds to SR.  The position of the resonance,
\eqt{\left(\Gamma_0'\right)_\textrm{SR}}, is found to be in good agreement with the position observed for the enhancement of spatial diffusion
(section~\ref{macroscopic}) and for the spatial distribution in moving frames (section~\ref{bunching}). We therefore conclude that the spectroscopy
methods provide a new means for observing SR on the propagation modes in a dissipative optical lattice.

\section{Conclusion}
\label{conclusion}
In conclusion, we have performed a detailed theoretical study of the influence of the noise related to the dissipative processes (optical pumping) on the
excitation process of the Brillouin propagation modes in an optical lattice. Our results are mainly obtained from semi-classical Monte-Carlo
simulations. We have shown that there is a specific amount of noise for which the excitation of the modes is resonant. This corresponds to the phenomenon
of stochastic resonance (SR), obtained when the Hamiltonian motion of an atom in the light potential and the dissipative processes in the lattice are precisely
synchronized. This result can be understood from the fact that the excitation of a mode involves synchronized half-oscillations in the potential wells and optical
pumping cycles. We have given a direct evidence for SR by means of the bunching properties in frames moving in the direction and at the same velocity as
the propagation modes, and an indirect evidence by the observation of the resonant enhancement of spatial diffusion. We have shown that the
resonant excitation of the modes, and thus SR, can be observed by means of pump-probe spectroscopy. The three methods give results for the dependence
of the SR point versus the lattice parameters in good agreement. We have also derived a simple analytical model for the synchronization condition in excellent
qualitative as well as quantitative agreement with the results of our numerical simulations.

The system that we examined in this paper is thus a good candidate for the experimental realization of SR in a system where the relevant parameters for
Hamiltonian motion and dissipative effects can vary independently and  that can be probed easily. Note also that optical lattices can be designed in a large
variety of configurations \cite{grynberg01}. This system may thus model many different physical situations.
First experimental results have been already obtained in a variant of the configuration considered in this paper \cite{stochastic2}. In Ref.~\cite{stochastic2},
Schiavoni {\it et al}
show that the excitation of the Brillouin propagation modes is nonmonotonic versus the noise strength. Our model-system is known to correctly simulate
the real experimental situations. We therefore expect our results to be reproductible in current experiments. In particular, it would be interesting to
investigate the atomic bunching in moving frames in order to show the coherent collective excitation aspect of the propagation modes. This may be
achieved {\it via} Bragg scattering, a technique that has been already used to give evidence of atomic localization in optical lattices
\cite{bragg95}. The characterization of SR {\it via} pump-probe spectroscopy experiments, a technique well-mastered in experiments on cold atoms
\cite{verkerk92,hemmerich93}, would also give interesting results. At last, an experimental investigation of the lattice parameters dependence of the
SR point (\eqt{\left(\Gamma_0'\right)_\textrm{SR}}) would give interesting insights onto the excitation process of the Brillouin propagation modes
in dissipative optical lattices.

\paragraph*{}
\acknowledgments
We are grateful to Anders Kastberg for useful comments on the paper. We also thank Ferruccio Renzoni and Michele Schiavoni for various discussions.
This work was supported by CNRS, the European Commission (TMR network ``Quantum Structures'', contract FMRX-CT96-0077) and
by R\'egion Ile de France under contract E.1220. Laboratoire Kastler Brossel is an ``unit\'e mixte de recherche de l'Ecole Normale Sup\'erieure
et de l'Universit\'e Pierre et Marie Curie associ\'ee au Centre National de la Recherche Scientifique (CNRS)''.


\end{document}